\documentclass{aa}

\usepackage{natbib}
\usepackage{graphicx}
\usepackage{txfonts}
\usepackage{multirow,array,makecell}
\usepackage{amsmath}
\usepackage{amssymb}
\usepackage{bm}
\def\degree{${}^{\circ}$}
\usepackage[colorlinks=true,linkcolor=blue,citecolor=blue,urlcolor=blue]{hyperref}
\usepackage{lineno}
\usepackage[normalem]{ulem}
\DeclareMathOperator{\lcm}{lcm}
\usepackage[flushleft]{threeparttable}
\usepackage{tabularx}

\begin{document}


   \title{Planetary climate under extremely high vertical diffusivity}

   \author{Yidongfang Si\inst{1,2}
          \and
          Jun Yang\inst{1*}
          \and
          Yonggang Liu\inst{1}}

   \institute{Department of Atmospheric and Oceanic Sciences, School of Physics, Peking University, Beijing 100871, China. *\email{junyang@pku.edu.cn}
   \and
   Now at the Department of Atmospheric and Oceanic Sciences, University of California, Los Angeles, CA 90095, USA
   }

 
  \abstract
   {}{Planets with large moon(s) or those in the habitable zone of low-mass stars may experience much stronger tidal force and tide-induced ocean mixing than that on Earth. Thus, the vertical diffusivity (or, more precisely, diapycnal diffusivity) on such planets, which represents the strength of vertical mixing in the ocean, would be greater than that on Earth.
   In this study, we explore the effects of extremely high diffusivity on the ocean circulation and surface climate of Earth-like planets in one asynchronous rotation orbit.}
   {The response of planetary climate to 10 and 100 times greater vertical diffusivity than that found on Earth is investigated using a fully coupled atmosphere--ocean general circulation model. In order to perform a clear comparison with the  climate of modern Earth, Earth's orbit, land--sea configuration, and present levels of greenhouse gases are included in the simulations.}
   {We find that a larger vertical diffusivity intensifies the meridional overturning circulation (MOC) in the ocean, which transports more heat to polar regions and melts sea ice there. Feedback associated with sea ice, clouds, and water vapor act to further amplify surface warming. When the vertical diffusivity is 10 (100) times the present-day value, the magnitude of MOC increases by $\approx$3 (18) times, and the global-mean surface temperature increases by $\approx$4$^{\circ}$C (10$^{\circ}$C). This study quantifies the climatic effect of an extremely strong vertical diffusivity and confirms an indirect link between planetary orbit, tidal mixing, ocean circulation, and surface climate. Our results suggest a moderate effect of varying vertical ocean mixing on planetary climate.
   }{}

   \keywords{tidal mixing -- vertical diffusivity --
             ocean circulation --
             planetary climate}

   \maketitle

\section{Introduction}\label{sec:introduction}

Tides, including atmospheric thermal tides, ocean tides, and solid tides, are important for the evolution of planetary orbital configuration and climate. For low-mass (K and M) star systems, previous studies found that the strong tidal torques caused by the host stars can significantly promote the processes of planetary orbit decay and circularization \citep{kasting1993habitable, barnes2009tidal, heller2011tidal, leconte2015asynchronous, barnes2017tidal, auclair-desrotour2018oceanic}. Meanwhile, heating through ocean tides and solid tides can influence the planetary surface energy flux and consequently atmospheric circulation and surface climate \citep{jackson2008tidal, barnes2013tidal, haqq2014geothermal, haqq2018exploring}. For instance, tidal heat flux on terrestrial planets with large eccentricities orbiting low-mass stars may reach hundreds of watts per meter squared \citep{barnes2013tidal}. This level of tidal heating is two orders of magnitude more than that on Jupiter's satellite Io and is high enough to trigger moist or runaway greenhouse effects, causing the planets to lose any surface water they may have \citep{barnes2009tidal,driscoll2015tidal,barnes2018gravitational}. These studies generally focused on the direct effect of tidal orbit or tidal heating. However, the indirect effect of ocean tides on planetary climate through oceanic mixing and meridional overturning circulation (MOC) has not been studied for exoplanets \citep{hu2014role, yang2014water, cullum2016importance, cael2017ocean, way2018climates, del2019habitable, yang2019ocean, yang2019seaice, Checlairetal2019, Olsonetal2020, zengandyang2021}. In this study, we focus on this problem, in particular for planets with much stronger tides than that on modern Earth.

On Earth, ocean tide is one of the main sources of mechanical energy that drives the deep oceanic MOC below the level of $\approx$2000~m \citep{munk1998abyssal,wunsch2004vertical,marshall2012closure}.
Assuming that the ocean is incompressible, a general expression to quantify the static stability of the ocean is the buoyancy frequency $N\,\equiv ({- \frac{g}{\rho_0} \frac{\partial \rho}{\partial z}})^{-\frac{1}{2}}$ (e.g., Eq. 2.224 in \citealp{vallis2017atmospheric}), where $g$ is the gravitational acceleration, $\rho$ is the locally referenced potential density, $\rho_0$ is the reference density, and $z$ is positive upward.
As the ocean has a stable stratification ($\rho$ increases with depth, $N^2>0$, e.g., \citealp{talley2011descriptive} and \citealp{king2012buoyancy}), the upwelling in the MOC requires a process capable of working against the negative buoyancy force and providing the potential energy. Through microstructure observations, \cite{waterhouse2014global} quantified the main resources for deep ocean mixing: about 1.5 TW (1 TW = 10\,$^{12}$ W) from tide-induced internal waves (``ocean tidal mixing''), 0.3 TW from wind-induced near-inertial waves, and 0.2 TW from internal lee waves\footnote{Lee waves are internal gravity waves generated when large-scale quasi-steady ocean flows encounter relatively small-scale bathymetric features, such as sea mountains (e.g., \citealp{bell1975lee}). Figure 1 of \cite{MacKinnon2013} is a nice visualization of lee waves in the ocean. The breaking of lee waves has been shown to be important for turbulent mixing at deep ocean \citep{nikurashin2013overturning,MacKinnon2013}.} \citep{nikurashin2013overturning}. These values have large uncertainties. On exoplanets, the strength of deep ocean mixing is unknown, and so it is necessary to investigate how the uncertainty on deep ocean mixing could influence planetary climate and habitability.

The strength of tidal mixing mainly depends on the intensity and frequency of tidal force, seawater stratification, ocean depth, and seafloor bathymetry \citep{stewart2008introduction,waterhouse2014global,auclair-desrotour2018oceanic}. For terrestrial planets in the habitable zone around low-mass stars, the tidal force should be orders of magnitude greater than that on Earth, because the habitable zone is very close to the host stars and the tidal force is inversely proportional to the cube of the orbital distance \citep{lingam2018implications}; for further details, please see Appendix A. Consider one low-mass star---planet system: if the planet is the size of Earth and is in the liquid-water habitable zone, such as $M_{\star}=0.3\,M_{sun}$ (stellar mass), $s\,=\,0.1$\,AU (distance between the star and the planet), and $R_p\,=\,R_{Earth}$ (planetary radius), the tidal force on the planet is approximately 300 times that on Earth from the Sun. If early Venus of our own Solar System had had an ocean, tidal dissipation could have varied by more than five orders of magnitude, which mainly depends on the rotation rate and ocean depth \citep{green2019consequences}.

For 1:1 tidally locked planets (alternatively called synchronously rotating planets, with orbital and rotation periods of equal length) with zero orbital eccentricity, tides would be steady and have no effect on ocean mixing. For asynchronously rotating planets, the effect of tidal mixing should be considered. The timescale for the evolution of one habitable planet from asynchronous rotation to synchronous rotation depends on many factors, such as orbital distance, stellar mass, energy dissipation rate, planetary rigidity, initial planetary rotation rate, land--sea configuration, and ocean bottom topography \citep{kasting1993habitable,barnes2017tidal}. Other factors can also influence the timescale and the final orbital state, such as thermal atmospheric tides, which have been suggested to explain the observed orbital resonance of Venus (\citealp[e.g.,][]{gold1969atmospheric,ingersoll1978venus}). \cite{leconte2015asynchronous} suggested that even with a relatively thin atmosphere, thermal atmospheric tides can drive Earth-like planets away from a synchronous rotation state. Moreover, planets with large eccentricities cannot be in 1:1 tidally locked orbit \citep{barnes2017tidal}.

Though previous studies have revealed that the change in tidal mixing (1.5--3 times the present level) has an impact on global MOC and paleoclimate \citep{schmittner2015glacial,weber2017tidal,wilmes2019glacial}, the effect of extremely vigorous tidal mixing (up to 100 times larger; see Appendix A) on the climates of exoplanets was not investigated.
In this study, we intend to address two questions through numerical experiments: (1) How does the MOC respond to extremely large tidal mixing? (2) How does the modified MOC influence the surface climate? We present the model description and experimental design in Sect. \ref{sec:methods}, describe the experimental results in Sect. \ref{sec:results}, and then summarize our key findings and discuss the broad implications in Sect. \ref{sec:summary}. \\

\section{Model description and experimental design}\label{sec:methods}

The model used in this study is the Community Climate System Model version 3 (CCSM3, \citealp{collins2006community}), which is a coupled climate model with four separate components and one central coupler to simultaneously simulate the atmosphere, ocean, land surface, and sea ice. The dynamical atmosphere component of CCSM3 is the Community Atmosphere Model version 3 (CAM3, \citealp{collins2004description}) with 96 longitudinal points, 48 latitudinal points (T31), and 26 vertical levels. The land surface model --- the Community Land Model version 3 (CLM3, \citealp{oleson2004technical}) --- has an identical horizontal grid to the atmosphere model. The ocean component of CCSM3 is the Parallel Ocean Program  (POP, \citealp{smith2004parallel}) version 1.4.1 with 116 latitudinal points, 100 longitudinal points, and 25 vertical levels in depth. POP's latitudinal resolution varies from $\approx$0.9 degrees near the equator to $\approx$3.0 degrees near the poles, and its layer thickness gradually increases from 12 meters near the sea surface to 450 meters near the ocean bottom. 
The sea-ice model of CCSM3 is the Community Sea-Ice Model version 5 (CSIM5, \citealp{briegleb2004scientific}), which includes both elastic-viscous-plastic dynamics and thermodynamics. The ocean and sea-ice models of CCSM3 share the same horizontal grid. Through modifying stellar spectrum, planetary orbit, atmospheric composition, surface pressure, surface topography, ocean bottom bathymetry, and land--sea distribution, CCSM3 has been employed to simulate the climates of different types of potentially habitable exoplanets \citep{yang2013stabilizing, hu2014role, yang2014water, yang2019ocean, yang2019seaice, zengandyang2021}.


\begin{table*}[!htbp]
\centering
\begin{threeparttable}
    \caption{\label{table:results}Global and one hundred-Earth-year mean characteristics of the simulated climate in three experiments with varying vertical diffusivity.}
        \begin{tabularx}{\textwidth}{p{1.2cm}p{1.2cm}p{1.2cm}p{1.2cm}p{1.2cm}p{1.2cm}p{2cm}p{2cm}p{2cm}p{2cm}}
            \hline
            \noalign{\smallskip}
            \hline
            \noalign{\smallskip}
          K$_t$ & T$_s$ & $\Phi^{o}_\mathrm{max}$ & $\Phi^{a}_\mathrm{max}$ & 
         $A_p$ & $A_s$ & SWCRE & LWCRE  & NCRE & TWV \\
          (K$_\mathrm{t0}$) & ($^\circ$C) & (Sv) & (Sv) & (0--1) & (0--1) & 
                   (W\,m$^{-2}$) & (W\,m$^{-2}$) & (W\,m$^{-2}$) &  (kg\,m$^{-2}$) \\                  
                     \noalign{\smallskip}
            \hline
            \noalign{\smallskip}
         1 & 14.0 & 18 & 100 & 0.32 & 0.12 & $-$55 & 28 & $-$27 & 22 \\
         10 & 18.2 & 60 & 90 & 0.30 & 0.11 & $-$53 & 30 & $-$23 & 26 \\
         100 & 23.5 & 324 & 40 & 0.28 & 0.09 & $-$48 & 30 & $-$18 & 37 \\
                     \noalign{\smallskip}
            \hline
            \noalign{\smallskip}
        \end{tabularx}
          \begin{tablenotes}
      \small
      \item \textbf{Notes.} T$_s$: surface air temperature. 
        $\Phi^{o}_\mathrm{max}$: maximum oceanic meridional overturning streamfunction 
        (1~Sv\,$=$\,10$^6$\,m$^3$\,s$^{-1}$, being approximately equal to 10$^9$\,kg\,s$^{-1}$ after considering seawater density). $\Phi^{a}_\mathrm{max}$: maximum value of the streamfunction in the Hadley Cells.
        $A_p$: planetary albedo. $A_s$: surface albedo. SWCRE: shortwave cloud radiative effect at the top of the atmosphere (TOA). LWCRE: longwave cloud radiative effect at TOA. NCRE: net cloud radiative effect (SWCRE plus LWCRE). TWV: vertically integrated water vapor amount in the atmosphere.
    \end{tablenotes}
  \end{threeparttable}
\end{table*}

\begin{figure*}[!htbp]        
 \center{\includegraphics[width=18cm]{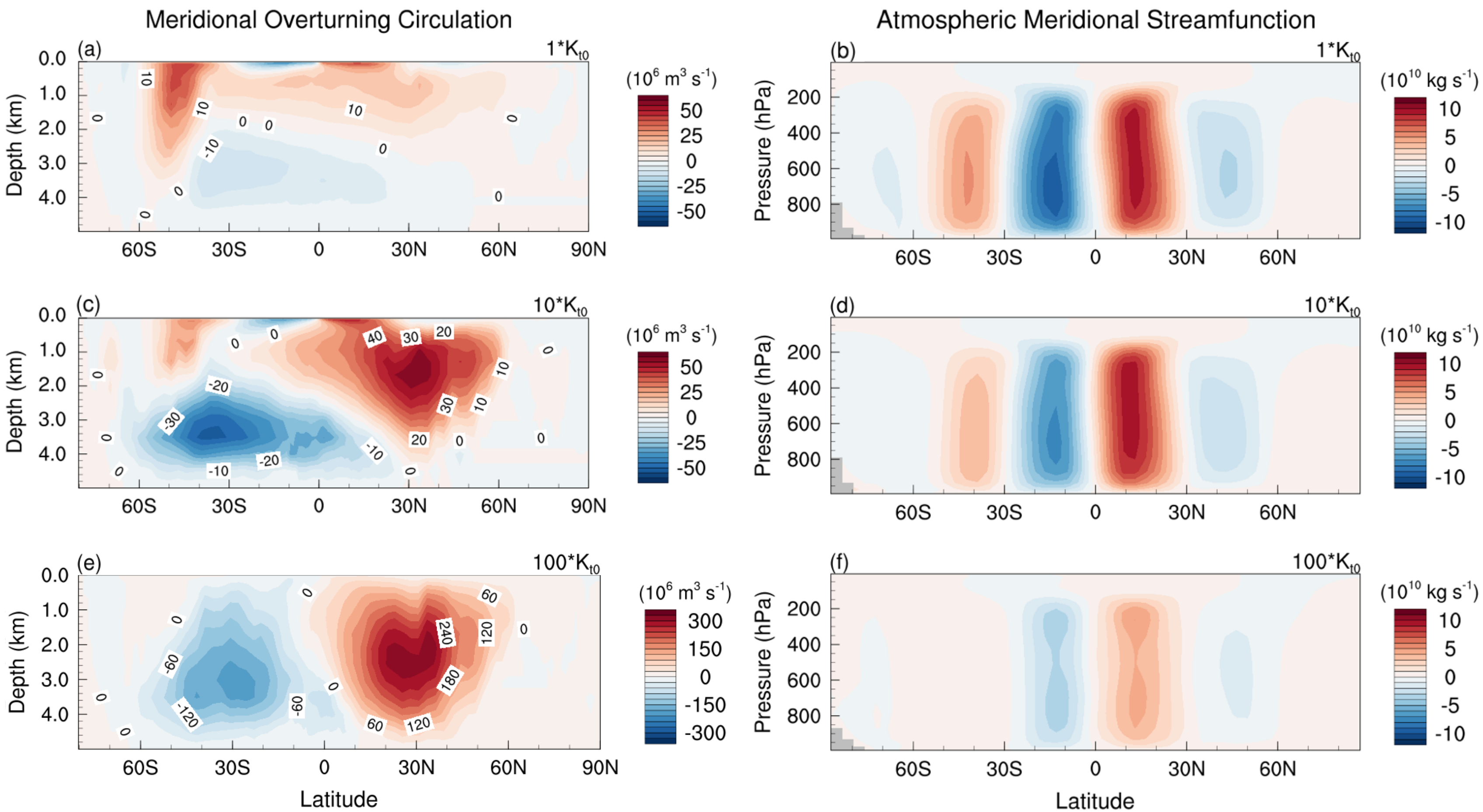}}        
 \caption{One hundred-Earth-year mean streamfunction of oceanic meridional overturning circulation (MOC, left column) and atmospheric meridional circulation (right column) in the experiments with the default diffusivity (a-b), and 10 times (c-d) and 100 times (e-f) the default diffusivity.
 The positive values represent the clockwise cells, and the negative values represent the anti-clockwise cells. The color bar in (e) is different from that in (a) and (c). 
 } 
 \label{streamfunction_1stGroup}
\end{figure*}

\begin{figure}[!htbp]        
\center{\includegraphics[width=9cm]{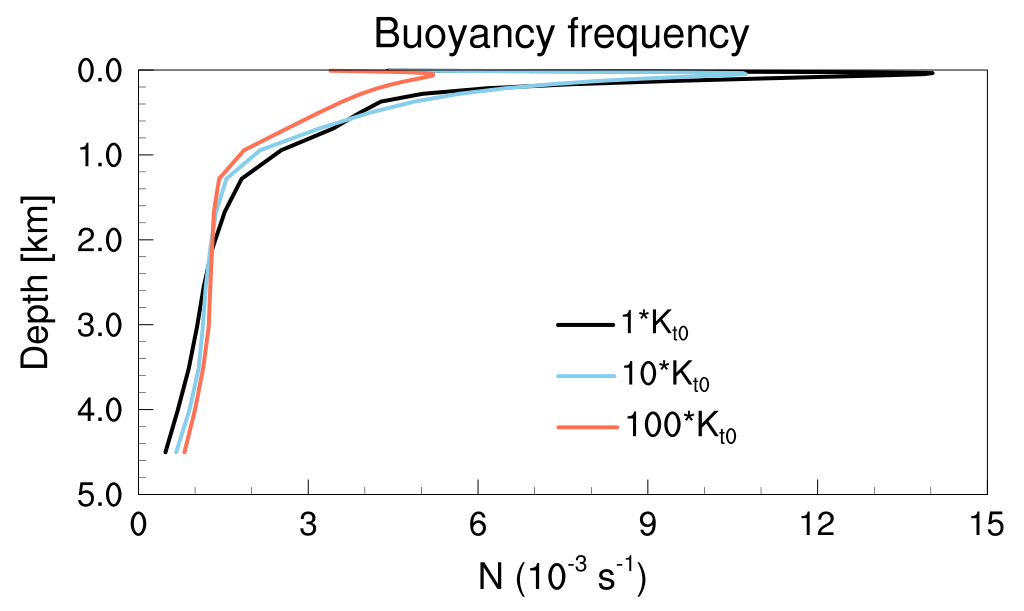}}        
\caption{Global and 100-Earth-year mean buoyancy frequency $N$\,$\equiv ({- \frac{g}{\rho_0} \frac{\partial \rho}{\partial z}})^{-\frac{1}{2}}$ of the ocean (e.g., Eq. 2.224 in \citealp{vallis2017atmospheric}) in three experiments, where $g$ is the gravitational acceleration, $\rho$ is the locally referenced potential density of the ocean, and $\rho_0$ is the reference density. Above the pycnocline, the surface ocean (above 1000~m) is strongly stratified, while in the deep ocean (below 2000~m) the stratification is very weak. The value of $N$ for the Earth’s ocean ranges from 10$^{-3}$ to 10$^{-2}$ s$^{-1}$ in the upper ocean, where density changes dramatically with depth. The deep and abyssal ocean is less stratified compared to the upper ocean, with $N$ ranging from 10$^{-3}$ to 10$^{-4}$ s$^{-1}$ \citep{talley2011descriptive,king2012buoyancy}.
 }
 \label{buoyancyN}      
\end{figure}

Ocean tides and tidal mixing are not explicitly resolved in the model (or in most state-of-the-art ocean circulation models), but the importance of tides for ocean circulation is achieved through their influence on vertical mixing \citep{large1994oceanic}. The mixing itself is due to the breaking of internal waves, which are generated mainly by the interaction between the tides and ocean bottom topography. The horizontal wavelength of the primary internal waves is determined by the aspect ratio of the topography, and the vertical wavelength of internal waves --- which are relevant for vertical mixing --- is proportional to the ratio of flow speed and stratification \citep{klymak_legg_pinkel_2010}. To resolve the internal waves, the horizontal and vertical resolutions of the models need to be about 100 m and 10 m in order to represent sharp topographies reasonably well \citep{jithin2020intensification}. Such resolution is clearly too high for global ocean models. Moreover, the tidally induced transient flows can be two orders of magnitude faster than the mean flow in the ocean (typically 1 cm\,s$^{-1}$ on Earth), which greatly limits the computational efficiency. The small-scale turbulence and nonlinearity associated with wave breaking and mixing cannot be explicitly resolved by the state-of-art global ocean models \citep{Lamb2014}, and this can be achieved through so-called direct numerical simulation (DNS; e.g., \citealp{salehipour_peltier_mashayek_2015}).

Four types of vertical diffusion processes are considered in the model CCSM3, including shear instability, double diffusion, boundary layer mixing, and internal mixing associated with internal waves and tides \citep{smith2004parallel}. The first three processes are parameterized based on state variables of the ocean, such as Richardson number, buoyancy frequency, and density ratio, which evolve as a function of time and space \citep{large1994oceanic}. The strength of the last process, internal mixing, is a function of ocean depth, and ranges from 1.0\,$\times$\,10$^{-5}$~m$^2$\,s$^{-1}$ in the upper ocean to 1.0\,$\times$\,10$^{-4}$~m$^2$\,s$^{-1}$ in the deep ocean (see Appendix B for the vertical profiles), and is horizontally uniform and unvarying with time. Internal mixing is enhanced in the deep ocean, which is mainly due to the effect of increased tidal mixing around ocean ridges and other topographic features.
Three levels of background vertical diffusivity, which represents the strength of the internal mixing ($K_t$), have been examined in this study: the default value of the oceans of present-day Earth (labeled as 1*$K_\mathrm{t0}$), and 10 times (10*$K_\mathrm{t0}$) and 100 times (100*$K_\mathrm{t0}$) this value. Due to the longtime integration of the ocean model and the limitation of computational resources, we did not perform experiments with weaker tidal mixing than that seen on Earth (although it is possible; e.g., if the ocean bottom were flat, or the moon were smaller or farther away from Earth), but the effect of weaker tidal mixing can be inferred from the results presented below. 
In our experiments, the background vertical viscosity of the ocean ($K_{\nu}$, also called momentum diffusivity) is ten times $K_t$ by default, and is modified simultaneously with $K_t$. This is because, over a broad range of temperature and pressure, the ratio between thermal diffusivity $K_t$ and kinematic viscosity $K_{\nu}$, namely the Prandtl number, is approximately constant\footnote{The value of the Prandtl number ranges from slightly below 1 to about 10 for the ocean and is supposed to be dependent on the ocean state in terms of Renolds number and Richardson number (e.g., \citealp{salehipour_peltier_mashayek_2015}). However, the turbulent Prandtl number is assumed to be a constant in most modern large-scale ocean models, although different values may be adopted in different models. For example, the Modular Ocean Model version 5 (MOM5, \citealp{Griffiesetal2014}) and the MIT General Circulation Model (MITgcm, \citealp{klymak_legg_pinkel_2010}) assume a turbulent Prandtl number of 1, while the Parallel Ocean Program (POP) assumes a value of 10 \citep{Danabasogluetal2012} based on field observations \citep{Petersetal1988}.}. 

\begin{figure*}[!htbp]        
 \center{\includegraphics[width=16cm]{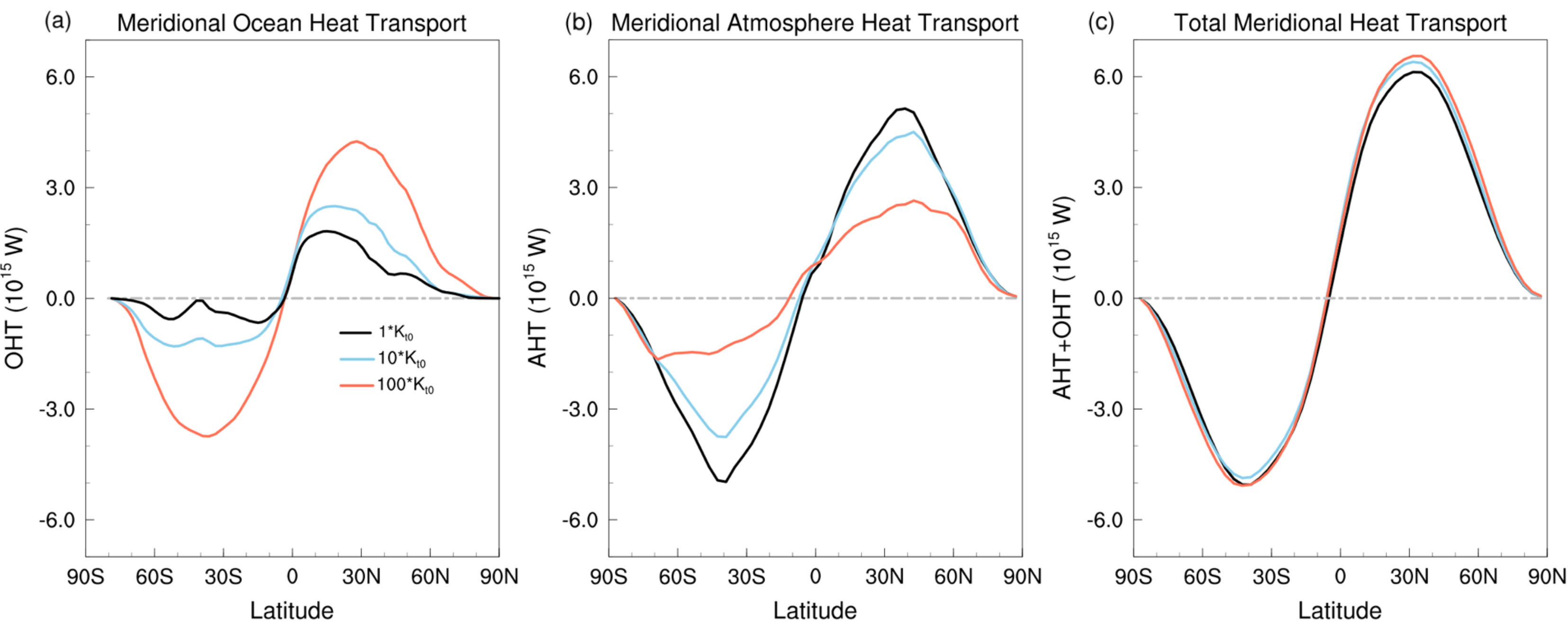}}        
 \caption{One hundred-Earth-year mean meridional (south-to-north) ocean heat transport (OHT, a), atmospheric heat transport (AHT, b), and total heat transport (c) in three experiments.} 
 \label{oht-aht-1stGroup}     
\end{figure*}

\begin{figure*}[!htbp]        
 \center{\includegraphics[width=16.5cm]{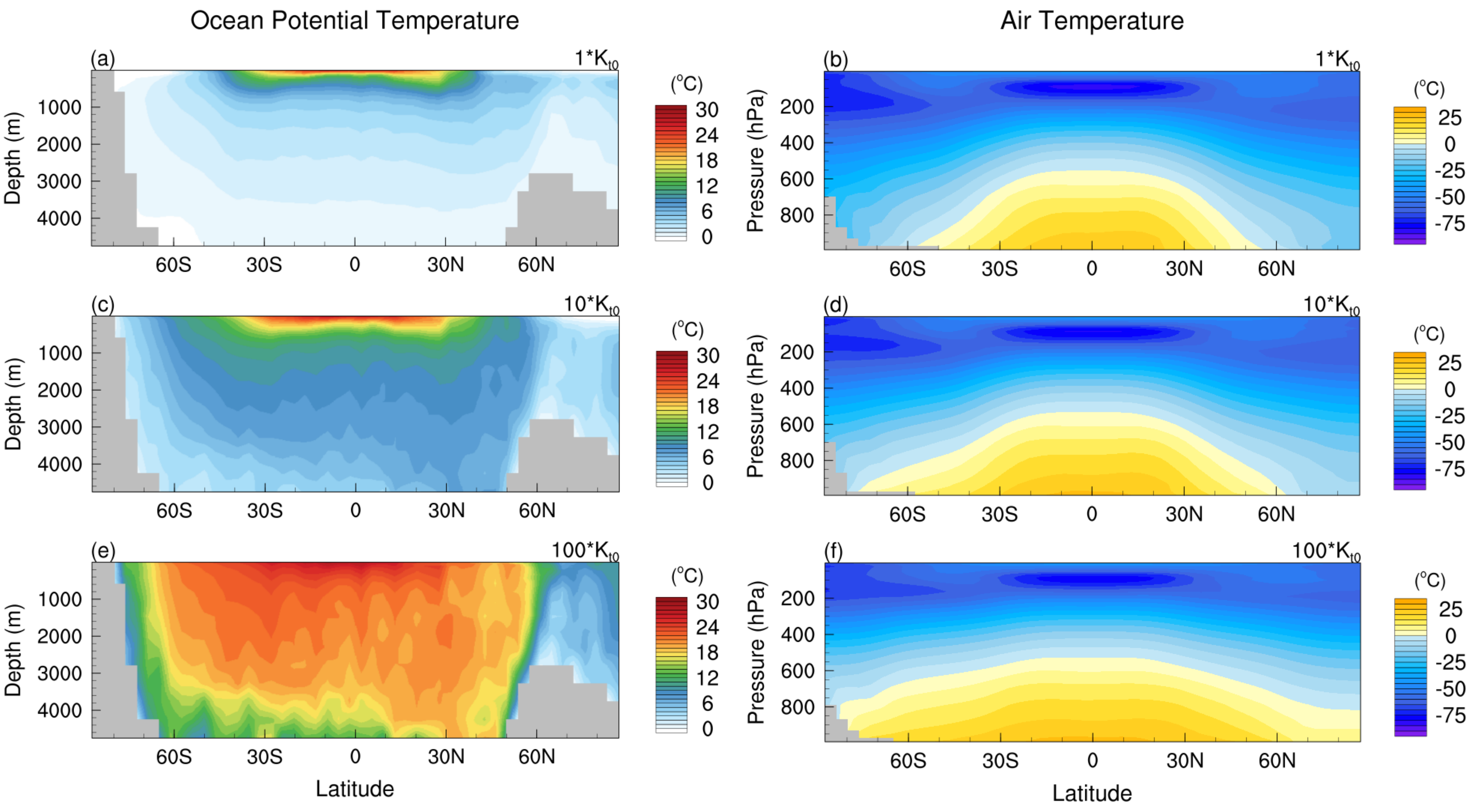}}        
 \caption{One hundred-Earth-year and zonal mean ocean potential temperature (left) and air temperature (right) in the experiments with the default diffusivity (a-b), and 10 times (c-d) and 100 times (e-f) the default diffusivity.} 
 \label{oceanT_airT_1stGroup}
\end{figure*}

In order to directly compare the simulation results with the climate of modern Earth, surface boundaries (including continental configuration, land surface topography, and ocean bottom topography) and parameters of the model are set to be the same as those of the Earth, except the vertical diffusivity. Planetary orbit is the same as that of Earth at present, where the orbital period is 365 Earth days, the rotation period is one day, the orbital obliquity is 23.44\degree, the orbital eccentricity is 0.0167, the planetary gravity is 9.8 m\,s$^{-2}$, the planetary radius is 6371 km, and the stellar flux at the substellar point is 1367 W\,m$^{-2}$. The atmospheric conditions in the simulations are also similar to those of the present Earth, that is, mainly composed of N$_2$ and O$_2$, and surface atmospheric pressure of 1.0 bar \citep[see][chapter 1]{hartmann2015global}. Concentrations of CO$_2$, CH$_4$, and N$_2$O are 355 parts per million volume (ppmv), 1714 parts per billion by volume (ppbv), and 311 ppbv, respectively.
Aerosol and ozone concentrations are the same as the pre-industrial levels. Each experiment is integrated for 1000 to 3000 Earth years to reach the equilibrium state, and the integration of the last 100 years is used to perform the analysis. \\


\section{Results} \label{sec:results}

\begin{figure*}[!htbp]        
 \center{\includegraphics[width=14.5cm]{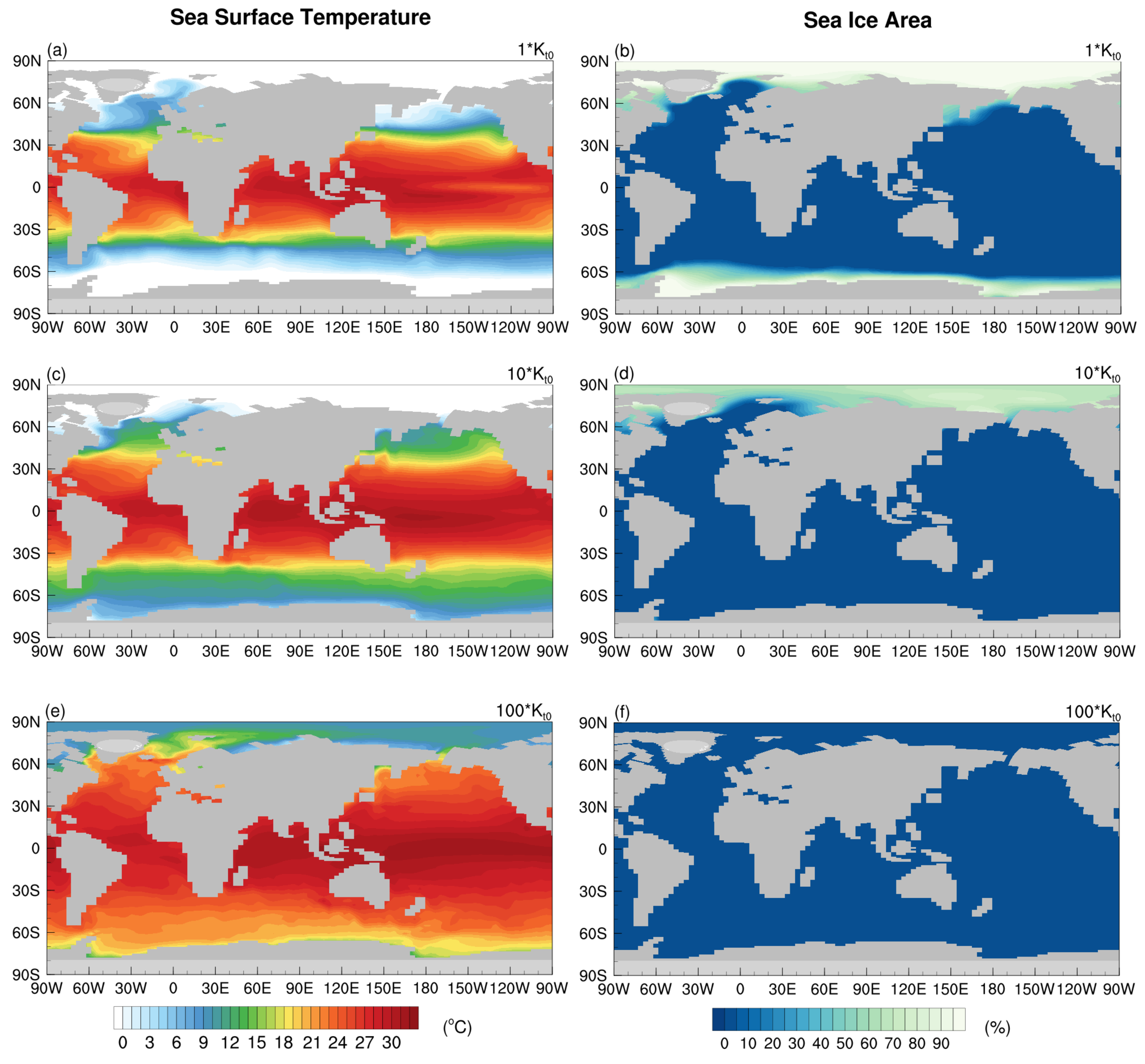}}        
 \caption{One hundred-Earth-year sea surface temperature (left column) and sea ice fraction (right column) in the experiments with the default diffusivity (a-b), and 10 times (c-d) and 100 times (e-f) the default diffusivity.} 
\label{sst-ice_1stGroup}
\end{figure*}

In this section, we answer the two questions introduced at the end of Sect. 1 by investigating the change of MOC and ocean heat transport with increased vertical diffusivity. We then describe the temperature change and the response of atmospheric circulation before discussing the three major feedback mechanisms that remarkably influence the global climate.

As shown in the left column of Fig.~\ref{streamfunction_1stGroup}, when the diffusivity is 100*$K_\mathrm{t0}$, the maximum value of the MOC increases from 18 to 324~Sv (1~Sv\,$=$\,10$^6$\,m$^3$\,s$^{-1}$), deep ocean overturning cells greatly strengthened. 
Here we utilize a simplified advective--diffusive balance to illustrate why the strength of MOC increases with greater vertical diffusivity. Consider a steady state with no source or sink of heat, the downward mixing of heat by diffusive processes below the thermocline is balanced by an upward advective heat flux, $W\partial_z\theta\approx \partial_z(K_t\partial_z\theta)$,
where $W$ is the mean vertical velocity and $\theta$ is the potential temperature
\citep{munk1966abyssal,vallis2000large,wunsch2004vertical, stewart2008introduction}. If the ocean stratification does not change significantly with $K_t$ (which is the case in our simulations; see Fig.~\ref{buoyancyN}), the MOC $\big($approximately proportional to $W \approx \partial_z K_t + K_t \partial_{zz} \theta / \partial_z\theta\big)$  increases significantly with larger $K_t$. More complicated scaling analyses using the advective--diffusive balance, thermal wind relation, mass continuity, and thermodynamic equation suggest that the intensity of MOC is approximately proportional to $K_t^{1/3}$ or $K_t^{2/3}$ \citep{welander1971thermocline, bryan1987parameter, vallis2000large, nikurashin2012a}. Our results are within these two ranges, with the MOC being about 3.3 (=\,10$^{0.52}$) and 18 (=\,100$^{0.63}$) times the reference MOC in 10*$K_\mathrm{t0}$ and 100*$K_\mathrm{t0}$ cases, respectively (Table~\ref{table:results}). 
As shown in Fig.~\ref{streamfunction_1stGroup}a, the MOC of the modern Earth is pole-to-pole. When $K_\mathrm{t}$ is very large (Fig.~\ref{streamfunction_1stGroup}e), all the deep water formed in the polar regions is brought up by the vertical mixing within its own hemisphere, and no Ekman suction is needed to bring up the water in the Southern Ocean as in the present day.

\begin{figure*}[!ht]        
 \center{\includegraphics[width=17cm]{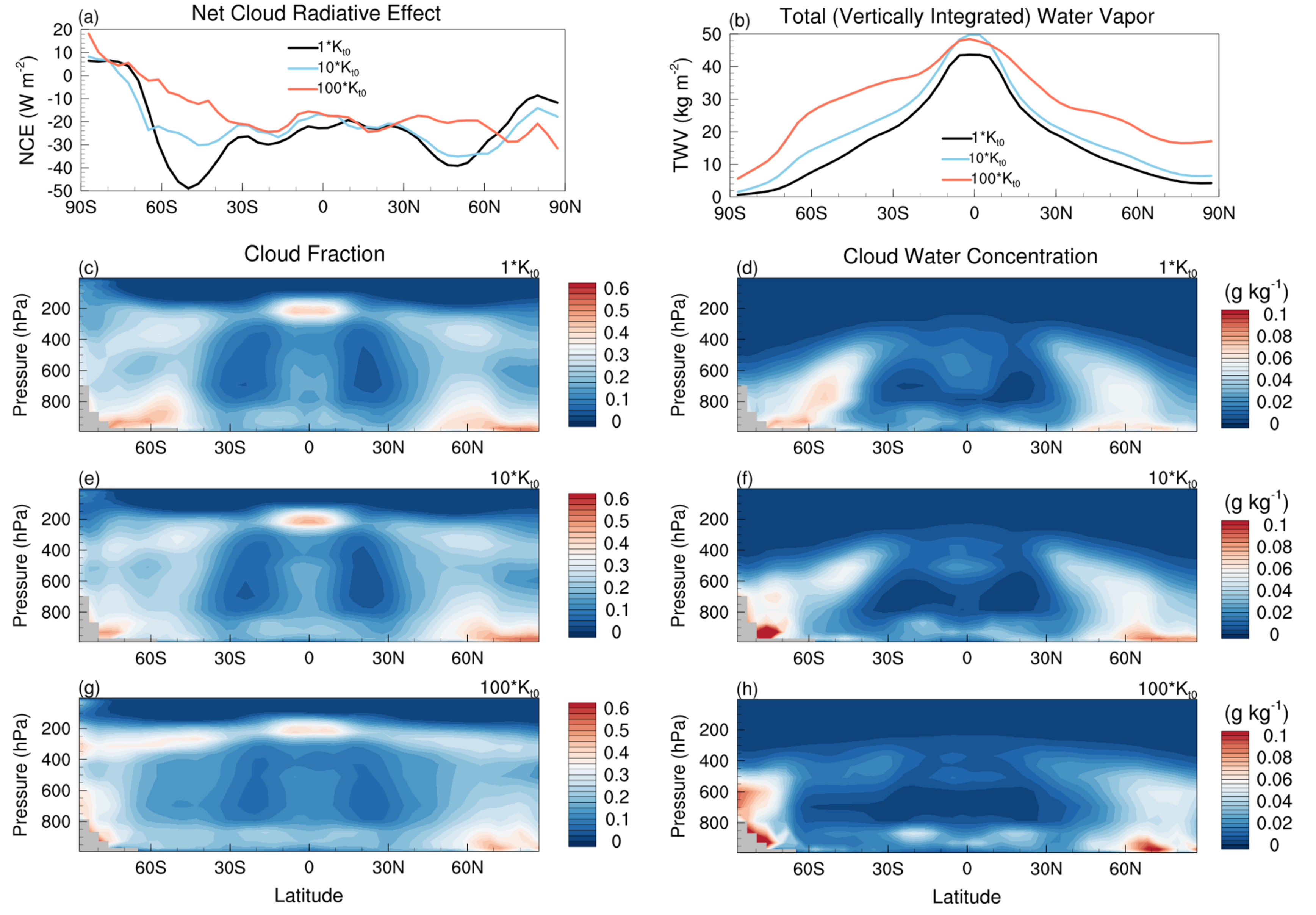}}
 \caption{One hundred-Earth-year and zonal mean net (shortwave plus longwave) cloud radiative effect at the top of the atmosphere (a), total (vertically integrated) water vapor amount in the atmosphere (b), cloud fraction (c, e, g), and cloud water concentration including liquid water and ice water (d, f, h) in three experiments with the default diffusivity (c-d), and 10 times (e-f) and 100 times (g-h) the default diffusivity.}  
 \label{cloud_1stGroup}    
\end{figure*}

Due to the enhanced MOC, the peak of the meridional ocean heat transport increases from 1.82 to 4.25~PW (1~PW $=$ 10$^{15}$~W) (Fig.~\ref{oht-aht-1stGroup}a). The ocean heat transport does not change proportionally with the MOC, because the MOC is substantially strengthened in the deep ocean, while the meridional ocean temperature gradient below the thermocline is much weaker than that of the surface ocean. 
This is consistent with previous studies of the vertical distribution of ocean heat transport on Earth: the shallow upper ocean circulation is more efficient in transporting heat because of the large temperature gradient there \citep{Boccalettietal2005,saenko2006vertical, ferrari2011processes}. 
Moreover, the latitude of the peak ocean heat transport shifts poleward by almost 20$^{\circ}$ in the 100*$K_\mathrm{t0}$ case  (Fig.~\ref{oht-aht-1stGroup}a). 
When $K_\mathrm{t}$ is relatively small, the ocean heat transport peaks at low latitudes because both the vertical temperature gradient of the ocean (Fig.~\ref{oceanT_airT_1stGroup}a) and the shallow meridional overturning driven by winds are large in the tropics.
In the case of very large $K_\mathrm{t}$, because of the substantial weakening of the wind-driven ocean circulation in the tropical and subtropical regions, and also the change in the structure of the MOC in the deep ocean (Fig.~\ref{streamfunction_1stGroup}a,c,e), the peak of the ocean heat transport now appears at 30$^{\circ}$ latitude where the streamfunction is the largest (Fig.~\ref{streamfunction_1stGroup}e).

In the simulations with 10*$K_\mathrm{t0}$ and 100*$K_\mathrm{t0}$, the planetary surface warms significantly (left column of Fig.~\ref{sst-ice_1stGroup}), reaching a global-mean surface temperature of 4.2$^{\circ}$C  and 9.5$^{\circ}$C above the reference temperature, respectively (Table~\ref{table:results}). Figure~\ref{oceanT_airT_1stGroup}e shows that the temperature of the deep ocean also increases significantly. With a large poleward ocean transport in the 100*$K_\mathrm{t0}$ case, the temperature of polar ocean surface rises to 10--20$^{\circ}$C (Fig.~\ref{sst-ice_1stGroup}e). As a result, the equator-to-pole surface temperature gradient reduces substantially. Moreover, given per unit energy, the change of surface temperature in the high latitudes is greater than that of the tropical regions, because the high-latitude surface is colder and loses less energy through thermal emission (the Planck effect, see \citealp{Lohmannetal2016}, chap. 12), and this also decreases the equator-to-pole temperature gradient.

The atmospheric heat transport includes three components: sensible heat transport, latent heat transport, and geopotential energy transport. The effects of both mean flows and different types of waves (such as Rossby waves and large-scale gravity waves, but not subgrid-scale gravity waves) are included in the calculation of atmospheric heat transport. The meridional ocean heat transport only includes the dominant term, namely the advective heat transport $\rho \mathrm{C}_p v \theta$, where $\rho$ is the seawater density, $\mathrm{C}_p$ is the specific heat capacity, $v$ is the meridional velocity, and $\theta$ is the potential temperature.
The predominant features of the atmospheric circulation response, including the tropical Hadley cells and the middle-latitude baroclinic eddies, become weaker with larger $K_t$ (see the right column of Fig.~\ref{streamfunction_1stGroup}) due to the decreased equator-to-pole temperature gradient. 
Consequently, the meridional heat transport carried by the atmosphere decreases (Fig.~\ref{oht-aht-1stGroup}b), compensating the increase in ocean heat transport. Figure~\ref{oht-aht-1stGroup}c shows clear compensation between atmospheric and oceanic heat transports, with total meridional heat transport increasing slightly with larger $K_t$ in the northern hemisphere. The compensation in the meridional heat transport between atmosphere and ocean is a common feature in past, present, and future Earth climates, known as the Bjerknes compensation \citep[e.g.,][]{Bjerknes1964, Stone1978, Yangetal2013, Yangetal2016}.

Above, we describe how the extremely large vertical mixing modifies the MOC, the surface temperature, and the atmospheric circulation. To better understand how the modified MOC influences the surface temperature, we investigated three essential feedback processes, including ice-albedo feedback, cloud feedback, and water vapor feedback (see \citealp{hartmann2015global}, chap. 10 and \citealp{Lohmannetal2016}, chap.12), as follows.

\textit{1) Ice-albedo feedback.}
The enhanced MOC transports more heat from the tropics to high latitudes (Fig.~\ref{oht-aht-1stGroup}a), melting the sea ice there. The planetary surface is able to absorb more stellar energy because the surface albedo decreases (Table~\ref{table:results}), which further melts the ice and snow in the polar regions. This positive ice-albedo feedback acts to substantially warm the surface and eventually melt all of the sea ice in the case with 100*$K_\mathrm{t0}$ (Fig.~\ref{sst-ice_1stGroup}f).

\begin{figure*}[!htbp]        
 \center{\includegraphics[width=13cm]{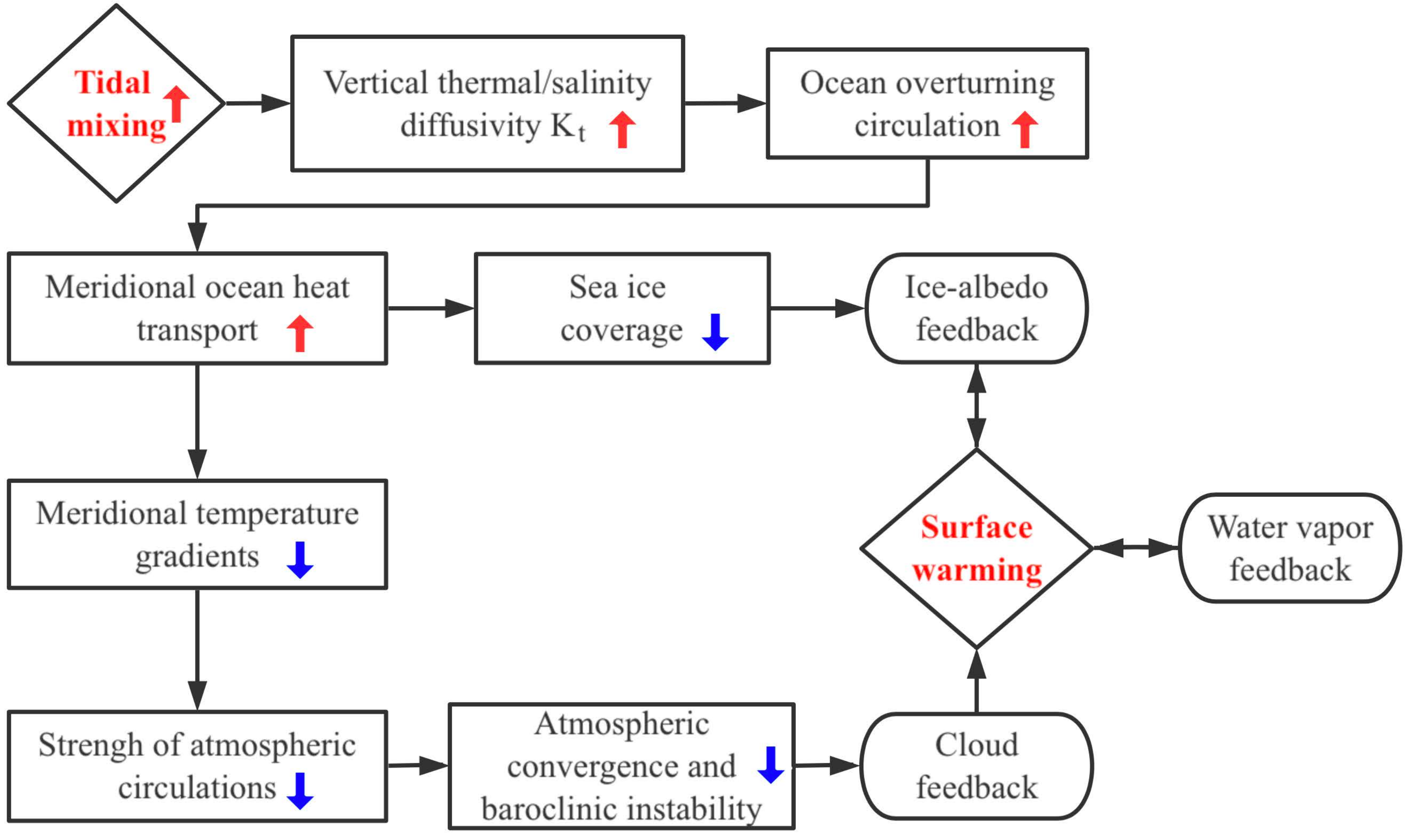}}        
 \caption{Schematic showing how stronger tidal mixing leads to global surface warming. The stronger tidal mixing enhances the oceanic MOC and thereby the meridional ocean heat transport. As a result, the polar sea ice coverage decreases, and the ice-albedo feedback warms the surface. The stronger ocean heat transport also decreases the meridional temperature gradient and weakens the atmospheric convergence and baroclinic instability of the atmosphere, and the cloud feedback further warms the surface. In addition, water vapor feedback also acts to warm the planetary surface.} 
\label{process_summary}
\end{figure*}

\textit{2) Cloud feedback.}
Fewer clouds form over the deep tropics with larger $K_t$ (Fig.~\ref{cloud_1stGroup}c,\,e,\,g), because the near-surface branches of the Hadley cells are weakened (right column in Fig.~\ref{streamfunction_1stGroup}), leading to less convergence and weaker convection in the intertropical convergence zone. 
At middle latitudes, fewer baroclinic eddies result in fewer stratoculumus clouds and layered clouds \citep{hartmann2015global}. 
At high latitudes, the clouds shift upward in altitude as the climate becomes warmer (Fig.~\ref{cloud_1stGroup}c,\,e,\,g). 
Moreover, the cloud water concentration (the ratio of cloud water mass to total air or cloud mass), which is a measure of the liquid and ice water amount in the atmosphere, decreases with larger $K_t$ (see Fig.~\ref{cloud_1stGroup}d,\,f,\,h).
The net result of these cloud responses is that the clouds become less reflective and the net cloud radiative effect (negative, a cooling effect) becomes smaller (Fig.~\ref{cloud_1stGroup}a). For modern Earth, the simulated global-mean net cloud radiative effect is $-$27 W\,m$^{-2}$, which is close to observations (see \citealp{hartmann2015global}, chap. 3). When the diffusivity is increased to 10 and 100 times the default value, the net cloud radiative effects are $-$23 W\,m$^{-2}$ and $-$18 W\,m$^{-2}$, respectively (Table~\ref{table:results}). Therefore, the cloud feedback also has a warming effect on the planetary surface with larger $K_t$.

\textit{3) Water vapor feedback.}
When the atmosphere becomes warmer, more water vapor can be maintained in the atmosphere following the Clausius-Clapeyron relation (e.g., \citealp{hartmann2015global}, chap. 1). As shown in Fig.~\ref{cloud_1stGroup}b, the vertically integrated water vapor amount in the atmosphere increases greatly with larger $K_t$. Acting as an effective greenhouse gas, the increased water vapor further warms the planetary surface.


\section{Summary and Discussion} \label{sec:summary}

In this paper, we present three coupled global atmosphere-ocean numerical experiments that demonstrate the effect of varying vertical diffusivity on the climate of planets in the habitable zone. The main findings of this study are summarized in Fig.~\ref{process_summary}. A stronger tidal mixing intensifies MOC in the ocean, transporting more heat poleward. The enhanced ocean heat transport melts the polar sea ice and reduces the surface albedo, increasing the sea surface temperature. Other feedback processes associated with clouds and water vapor act to further warm the global surface. This study confirms a connection between planetary orbit, tidal mixing, ocean circulation, and planetary climate. The change of MOC due to larger vertical diffusivity could further influence the partition of CO$_2$ between the atmosphere and ocean, as well as the spatial distributions of O$_2$ and nutrients in the ocean. 

The trend that we find, namely warming surface temperature with larger vertical diffusivity, and the responses of atmospheric circulation, clouds, and water vapor to increased oceanic heat transport are qualitatively consistent with previous studies of Earth's climate in the past and future \citep{rind1991increased, herweijer2005ocean, barreiro2011climate, rose2012ocean, koll2013tropical, knietzsch2015impact, rencurrel2018exploring, hilgenbrink2018response}. 
However, in previous investigations, the change in vertical diffusivity was much smaller than that employed here and the magnitude of surface warming was also smaller.

The conclusion of this study is based on the simulations using Earth's orbit and land--sea distribution. Further studies are required for planets with different continents. For Earth, strong tidal mixing is concentrated at the regions close to the ocean ridges \citep[e.g.,][]{waterhouse2014global}, and so the effect of tidal mixing on the planetary climate should also depend on the height, roughness, and spatial distribution of the topographic features of the seafloor. If the bottom of the ocean were flat, the tidal mixing would be weaker. Additionally, if the ocean were so deep that near-bottom ocean mixing were not able to influence the upper ocean, the effect of tide-induced warming would also be smaller than that shown in this paper. The vertical diffusivity in this work is fixed in time and is assumed to be horizontally uniform. Future work is required to employ an ocean tide model \citep[e.g.,][]{st2002estimating, egbert2004numerical, green2019consequences} to calculate the spatial pattern of the oceanic tides on different types of potentially habitable exoplanets, and to use high-resolution ocean general circulation models \citep[e.g.,][]{jansen2016turbulent} to resolve the strength and spatial pattern of tidal mixing. The present computational speed and resources do not allow for this kind of simulation with a global-scale domain. Moreover, the effect of tidal heating, which can directly heat the ocean bottom \citep{mashayek2013role}, has not been considered in this work and should be included in the future.\\

\begin{acknowledgements}
The data and code used in this article are available via: 
\href{http://doi.org/10.5281/zenodo.5527297}{http://doi.org/10.5281/zenodo.5527297}. J.Y. is supported by the National Natural Science Foundation of China under the grants 41888101 and 42075046. Y. Si acknowledges the scholarship provided by China Scholarship Council that supports her study at UCLA. The authors thank Mingyu Yan for her help in preparing the manuscript. The authors sincerely appreciate the reviewer and editors for their insightful comments and suggestions, which have greatly improved the manuscript.\\
\end{acknowledgements}

\bibliographystyle{aa}
\bibliography{ana_tide}

\begin{appendix} 

\section{Estimating the tidal mixing strength on potentially habitable planets around low-mass stars}

In this Appendix, we roughly estimate the background vertical diffusivity $K_t$ on potentially habitable planets around low-mass stars based on an analytic estimation of energy flux from ocean tides to internal waves \citep{bell1975lee} and the equilibrium tide theory. 

\subsection*{A.1. The energy flux from ocean tides to internal waves}
First, we introduce the energy flux from ocean tides to internal waves, which can be used to quantify the strength of ocean tidal mixing.
Theoretically, this energy flux can be estimated as \citep{bell1975lee,llewellyn2002conversion}:
\begin{equation}
C_\mathrm{Bell} =\rho_0 U_0^2 L N \sqrt{1-\frac{f_0^2}{\omega_0^2}} \int_0^\infty k\tilde{h}(k) \tilde{h}^\star(k) \frac{dk}{2\pi} \ 
\label{C_Bell}
\end{equation}  
\noindent in units of Watts, where $\rho_0$, $U_0$, $N$, and $f_0$ are the mean density of seawater, the amplitude of the barotropic tidal current, the buoyancy frequency, and the Coriolis frequency, respectively, $\omega_0$ is the fundamental tidal frequency, which we describe in Appendix A.3, $L$ denotes the horizontal scale of the bottom topography, $\tilde{h}(k)$ is the Fourier transform of the topography, $k$ is the wavenumber of the topography, and the star symbol denotes the complex conjugate. This equation is 
based on three assumptions: the fluid is unlimited in the vertical direction; the topography varies only in 
one direction; the buoyancy frequency is uniform in the ocean \citep{bell1975lee, llewellyn2002conversion, nycander2005generation}. 
In the following sections, we provide our scalings of $U_0$ and $\omega_0$ in order to estimate $C_\mathrm{Bell}$ and thus the background vertical diffusivity $K_t$ on the potentially habitable exoplanets.

\subsection*{A.2. The maximum tidal height and barotropic tidal current amplitude}
In this section, we estimate the tide-generating force and maximum tidal height, and then estimate the amplitude of the barotropic tidal current from the conversion between the potential energy and kinetic energy. In the equilibrium tide theory, the tide-generating potential is: 
\begin{equation} 
V=-\frac{GM_* R_p^2}{2s^3}(3\cos^2\theta -1), 
\label{tidal_potential}
\end{equation}
where $G$ is the gravitational constant, $M_*$ is the stellar mass, 
$R_p$ is the planetary radius, $s$ is the orbital distance, and $\theta$ is approximately equal to the zenith angle of the star \citep{macdonald1964tidal}. 
The horizontal component of the tide-generating force is: 
\begin{equation}
F_\parallel= -\frac{1}{R_p} \frac{\partial V}{\partial \theta} = - \frac{3GM_* R_p}{2s^3}\sin2\theta.
\label{tidal_force_2}
\end{equation}
The maximum tidal height \citep{kantha2000numerical,lv2012} is expressed as:
\begin{equation}
\eta_\mathrm{max}=\frac{1}{2} \frac{M_*}{M_p} \frac{R_p^4}{s^3} (3\cos^2\theta-1),
\label{max_tidal_height}
\end{equation}  
with the spherical triangle $\theta$ defined as
\begin{equation}
\cos\theta = \sin\delta \ \sin\varphi_0 + \cos\delta \ \cos\varphi_0 \ \cos T_1, 
\label{spherical_triangle}
\end{equation}
where $\delta$ is the declination, 
$\varphi_0$ is the latitude of the observer, and $T_1$ is the hour angle. For an M dwarf--exoplanet system with a typical stellar mass $M_*=0.3\,M_\mathrm{sun}$, typical orbital distance of planet in the habitable zone $s=0.1\ \mathrm{AU}$, Earth-like planetary radius $R_p =R_\mathrm{Earth}$, and planetary mass $M_p =M_\mathrm{Earth}$, the horizontal tidal force $F_\parallel$ and maximum tidal height $\eta_\mathrm{max}$ are about 300 times those on Earth from the Sun (Eqs.~\ref{tidal_force_2} and \ref{max_tidal_height}), and 100 times those on Earth contributed by both the Sun and the Moon.

Tidal energy contains kinetic energy in tidal currents, and potential energy due to the rise and fall of water levels. During a tidal cycle, the maximum tidal potential energy is $mg\eta_\mathrm{max}$, where $m$ is the mass of seawater and $g$ is the gravitational acceleration. The total kinetic energy of the tidal current can be scaled as the kinetic energy of the barotropic tides $mU_0^2/2$, because most of the tidal energy comes from the barotropic tides \citep{carter2008energetics}, the velocity of which is vertically uniform. Assuming that all the tidal potential energy is eventually transferred into kinetic energy, the squared barotropic tidal current amplitude ($U^2_0$) is proportional to the maximum tidal height ($\eta_\mathrm{max}$). From the aforementioned discussion, $U^2_0$ in Eq.~\ref{C_Bell} on such an exoplanet is about 100 times that on Earth.

\subsection*{A.3. The fundamental tidal frequency}
In this section, we provide the method to scale $\omega_0$, the fundamental tidal frequency in Eq.~\ref{C_Bell}. Fundamental frequency refers to the lowest frequency of oscillatory flows, which is typically the frequency of external forcing. Waves are generated at the fundamental frequency, as well as at higher harmonics. In quasistatic fluid, the energy conversion from linear barotropic tides to internal waves is predominantly into the fundamental tidal frequency $\omega_0$ \citep{llewellyn2002conversion,bell1975lee}. For barotropic tides,
\begin{equation}
\omega_0 = \frac{2\pi}{{T_{\eta}}_\mathrm{max}} \ ,
\label{def_omega_0}
\end{equation}
where ${T_{\eta}}_\mathrm{max}$ is the period of the maximum tidal height (e.g., Table 17.1 of \citealp{stewart2008introduction}), 
and is the same as the period of the tide-generating potential \citep{doodson1921harmonic, stewart2008introduction}. In Eq.~\ref{spherical_triangle}, the latitude of the observer $\varphi_0$ is constant, and the declination $\delta$ changes very slowly over time. Therefore, the period of the spherical triangle $\theta$ is determined by the period of the hour angle $T_1$, which is the least common multiple of $T_r$ and $T_o$ (denoted as $\lcm\{T_r, T_o\}$) for slowly rotating planets. We therefore have $\omega_0\approx 4\pi/\lcm\{T_r, T_o\}$, based on Eqs.~\ref{max_tidal_height}, \ref{spherical_triangle}, and \ref{def_omega_0}.

The Coriolis frequency at the latitude $\varphi_0$ is 
\begin{equation}
f_0 = 2\Omega \sin\varphi_0 =\frac{4\pi}{T_r} \sin\varphi_0,
\label{def_Coriolis}
\end{equation}
where $T_r$ is the rotation period of the planet. In middle latitudes where $\varphi_0 \approx 1/2$, $f_0 \approx 2\pi/T_r$.

Therefore, the term $\sqrt{1-f_0^2/\omega_0^2}$ in Eq.~\ref{C_Bell} is approximately $\sqrt{1-(\lcm\{T_r,T_o\}/T_r)^2/4}$ for a slowly rotating exoplanet. When $f_0/\omega_0\approx(\lcm\{T_r,T_o\}/T_r)^2/4\geq 1$, higher harmonics of tidal frequency (e.g., $2\omega_0$, $3\omega_0$,...) need to be considered. Here, we make a rough assumption that in Eq.~\ref{C_Bell}, the term $\sqrt{1-f_0^2/\omega_0^2}$ is within the same order of magnitude as that of the present-day Earth. (For the present-day Earth--Moon system, the frequency of M2 tide is about two times the Coriolis frequency on Earth, i.e., $\sqrt{1-f_0^2/\omega_0^2}\approx \sqrt{3/4}$.)

\subsection*{A.4 Compare $K_t$ and $C_\mathrm{Bell}$ on Earth and exoplanets}

To estimate the magnitude of $C_\mathrm{Bell}$ on the asynchronously rotating planets around low-mass dwarfs, we assume that $\rho_0$, $L$, and the integration related to the bottom topography in Eq.~\ref{C_Bell} remain the same orders of magnitude as that on Earth. 
Based on the analysis and assumptions in sections A.2 and A.3, on such an exoplanet, $U_0^2$ is around 100 times that on Earth, and $\sqrt{1-f_0^2/\omega_0^2}$ does not change significantly for slowly rotating exoplanets.
We further assume that in Eq.~\ref{C_Bell} the buoyancy frequency on such an exoplanet is similar to that on Earth. In our experiments with varying background vertical diffusivities, the global mean buoyancy frequency remains within the same order of magnitude (Fig.~\ref{buoyancyN}).

Therefore, the energy flux from tides to internal waves on such an exoplanet is approximately 100 times that on Earth. With the assumption that there is a linear relation between K$_t$ and C$_\mathrm{Bell}$ \citep{nycander2005generation}, the oceanic background vertical diffusivity K$_t$ on rocky planets in the habitable zone of low-mass stars may be 100 times that on Earth.

This estimate has a lot of caveats because there are so many unknowns of exoplanets. It oversimplifies many factors that could make a difference to the scaling, such as bottom topography, spherical triangle, orbital period, rotation period, and different ocean stratification. A more accurate way to estimate the strength of tidal mixing is to employ a numerical tidal model such as that used in \cite{green2019consequences}, but it would be much more expensive.

\section{Vertical profiles of background diffusivity}

\begin{figure}[!htbp]
\centering
{\includegraphics[width=9cm]{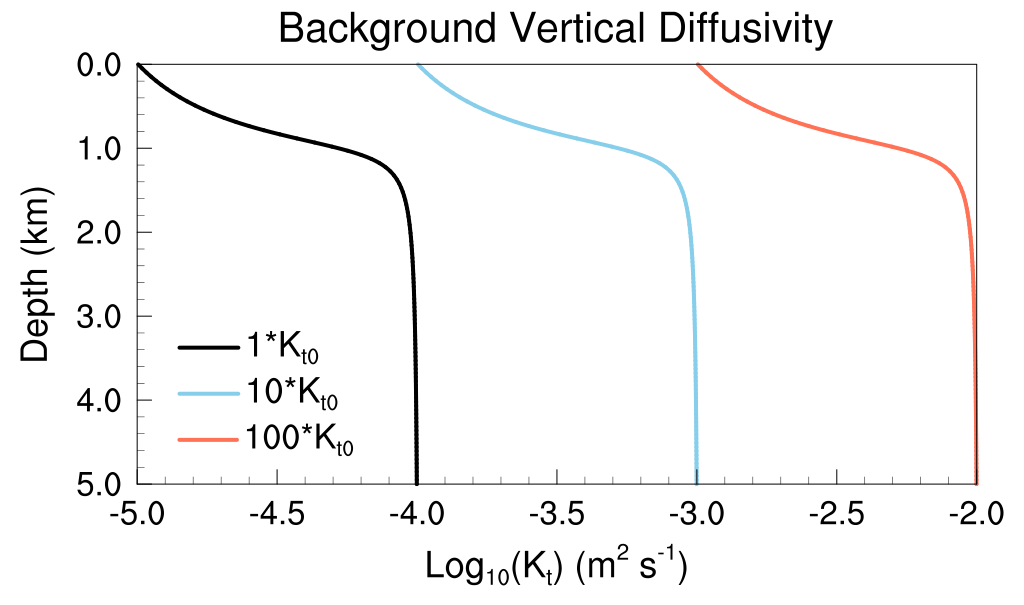}}
\caption{Profiles of the vertical thermal/salinity diffusivity K$_t$ 
in the model CCSM3. The black line denotes default diffusivity K$_\mathrm{t0}$, the blue line denotes 10 times  
the default diffusivity, and the red line denotes 100 times the default diffusivity. The ocean depth is the absolute value of $z$ in Eq.~\ref{Kt}.}
\label{Kt-depth}
\end{figure}

In this work, the key parameter is the background vertical thermal/salinity diffusivity in the ocean model. The K-Profile Parameterization \citep{large1994oceanic} is used to parameterize the unresolved vertical mixing. The background vertical thermal/salinity diffusivity ($K_t$) is
\begin{equation} 
K_t = \nu_1 - \nu_2 \times \arctan{\frac{z+H}{L}} \ ,
\label{Kt}
\end{equation}
\noindent where $\nu_1 = 5.24 \times 10^{-5}$ m$^2$\,s$^{-1}$ and $\nu_2 = 3.13 \times 10^{-5}$ m$^2$\,s$^{-1}$ by default, $z$ is the model vertical coordinate ($z$ is positive upwards, and $z=0$ for the mean sea level), $H = 1000$ m is the depth at which background vertical diffusivity is $v_1$, $L$ is the length scale where the transition in K$_t$ takes place and $1/L = 4.5 \times 10^{-3}$ m$^{-1}$ \citep{smith2004parallel}. Figure~\ref{Kt-depth} shows the three vertical profiles of K$_{t}$ employed in this study. The parameterizations of other types of mixing, including shear instability, convective instability, double diffusion, and boundary mixing, were not modified in any of the experiments.\\

\end{appendix}


\end{document}